\documentclass[12pt]{article}
\usepackage{arxiv}
\usepackage[utf8]{inputenc} 
\usepackage[T1]{fontenc}    
\usepackage{hyperref}       
\usepackage{url}            
\usepackage{booktabs}       
\usepackage{amsfonts}       
\usepackage{nicefrac}       
\usepackage{microtype}      

\usepackage{graphicx}      
\usepackage{cite}
\usepackage{algorithm}
\usepackage{subfloat}
\usepackage{amsmath}
\usepackage{float}
\usepackage[linesnumbered,ruled,vlined, algo2e]{algorithm2e}
\usepackage{algorithmicx}
\usepackage{chngcntr}
\usepackage{xcolor,soul,framed} 
\title{Anytime Planning: A Motion Planner for Dynamic Environment}

\author{
  Trishant ~Roy \\
  Department of Energy Sciences\\
  IIT Bombay\\
  Mumbai-400076, India \\
  \texttt{trishantroy@iitb.ac.in} \\
   \And
 Anindya ~Harchowdhury \\
  IITB-Monash Research Academy\\
  Mumbai-400076, India \\
  \texttt{anindya.harchowdhury@monash.edu} \\
     \And
  Leena ~Vachhani \\
  Systems and Control Engineering\\
  IIT Bombay\\
  Mumbai-400076, India \\
  \texttt{leena.vachhani@iitb.ac.in} \\
}

\begin{document}
\maketitle

\begin{abstract}
Motion planning in the presence of multiple dynamic obstacles is an important research problem from the perspective of autonomous vehicles as well as space-constrained multi-robot work environment. In this paper, we address the motion planning problem for multiple dynamic obstacle rich environment and propose a probabilistically, complete novel motion planning algorithm. Our claim is that given a fixed path cost i.e. the Euclidean path length, the proposed algorithm plans a path with the least computational time as compared to the state-of-the-art techniques. At the same time, given the time duration for planning, it plans the minimum cost path. Dynamic constraints have been taken into consideration while designing the planner such that the optimal planned path is feasible for implementation. The results of extensive simulation experiments show that the proposed sequential BIT* outperforms the DOVS both in the planned path length as well as the path generation time.   
\end{abstract}


\section{Introduction}
    Autonomous robotic systems have grown to become well-established research directions over the last few decades. The scope of applications in such self-driven systems is ever-increasing and has become one of the fastest-growing research areas in the present decade.  Either finding a collision freeway for an autonomous vehicle or training a mobile manipulator to move and perform tasks without colliding with static or dynamic objects falls in the category of path planning. It has been one of the cornerstone areas of investigation under the hood of robotics research. Finding a feasible, collision-free, and minimum energy path in a space-constrained static environment is always a challenging task. Safe robot navigation in a hybrid (static and dynamic) environment is even more challenging due to environmental uncertainty. It can pose a critical challenge for a perception dependent robot due to low confidence at finding good localization of the robot. This can take the robot even further away from achieving the primary goal to follow an optimal path.

Earlier methods such as reactive planning, and static planners have become more or less obsolete as they fail to address the need to produce a collision-free path faster by choosing the suitable constraint that narrows down the domain of search. Motion planning in dynamic environments poses a higher degree of challenges as it requires the robot to predict the future states of an obstacle within its Field of View (FoV). Predictive techniques help the robot plan a safe path and maneuver along the planned direction, which must accompany with kino-dynamic constraints of the robot to make it feasible. In this paper, we propose a novel motion planning algorithm in that a controller has been designed keeping in view of the dynamic constraints of the robot. This works on top of the novel path planning algorithm. The main contribution of the paper lies in the design of a motion planning algorithm that plans a maneuverable collision-free path with the minimum cost on the one hand, and computes a collision-free path with a given cost at the minimum time on the other, in comparison with other state-of-the-art path planning algorithms. To the best of our knowledge, this is the first research in which, {\em safety} as well as {\em path length} has been considered as the parameters of importance, and it not only takes care of the global planning, it takes account of local planning as well.

The rest of the paper is organized as follows. A thorough study on the classical as well as state-of-the-art path planning algorithms have been presented in Section 2. The problem statement follows it, in Section 3. Section 4 presents the proposed method, and its effectiveness is discussed then. Experimental set-up, along with implementation details, have been discussed in Section 5. Results based on the performance metrics chosen for the comparison, along with critical comments, have been stated in Section 6, and a summary of the motive and important findings of the work are included as a conclusion in Section 7.

\section{Related Works}
\begin{figure}[t]
	\begin{center}
		\includegraphics[width=8.4cm]{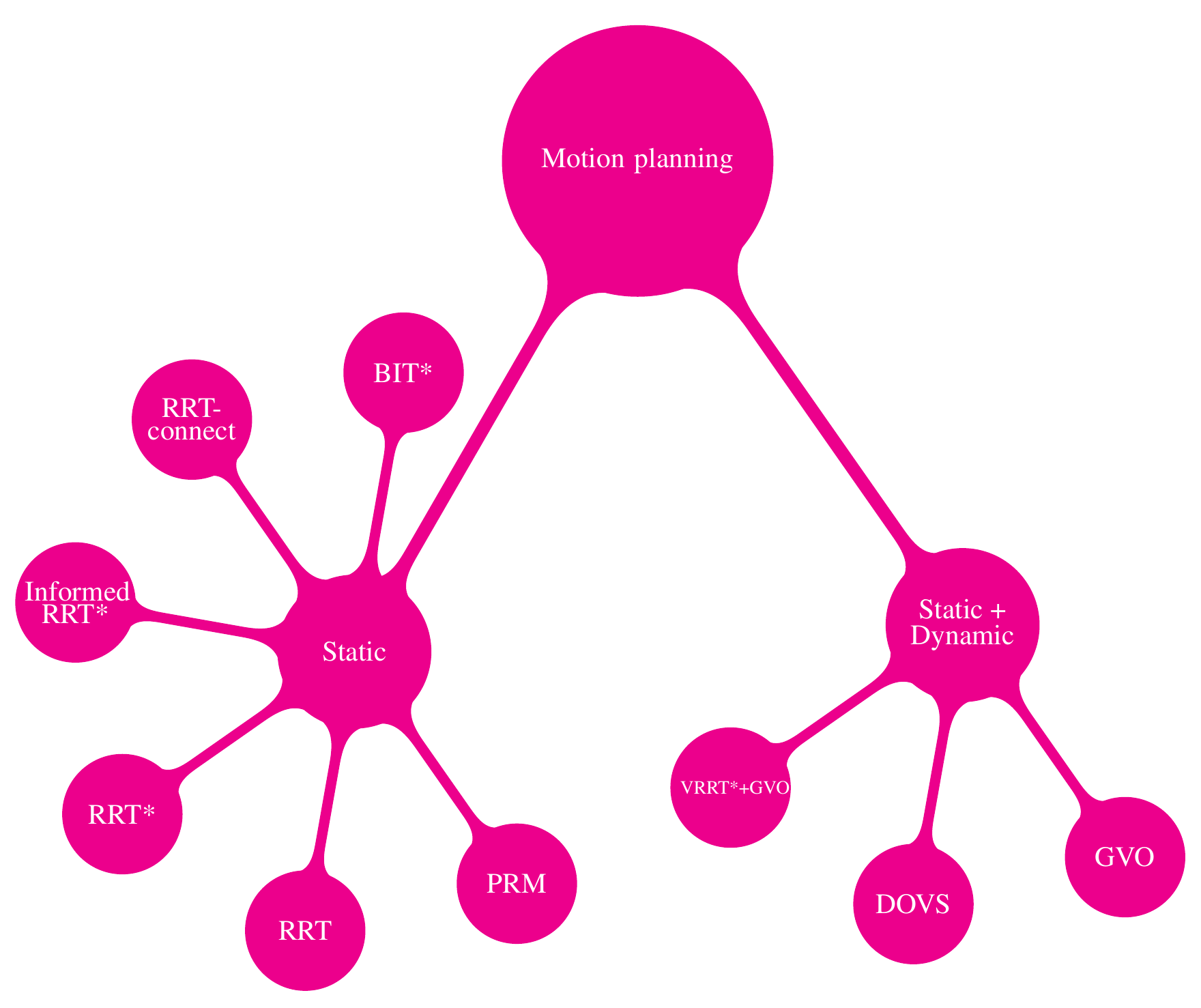}    
		\caption{State-of-the-art techniques for motion planning in an environment with static and/or dynamic objects.} 
		\label{fig:01}
	\end{center}
\end{figure}

    There is a rich legacy of research on motion planning for Autonomous Ground Vehicles (AGV). Many times, even high dimensional planners that are not designed for an application to navigate an autonomous vehicle are found to be very effective for path planning in 2D or 3D environments. For better understanding the goal of the research carried out in this paper, we first discuss the state-of-the-art in this area that closely associates with the objective of motion planning in static or dynamic environments. A map of these works is shown in Figure~\ref{fig:01}. 

A simple single-query path planning mechanism called RRT-Connect found in \cite{icra20}, combines Rapidly Exploring Random Tree (RRT) with a simple greedy heuristic that aggressively tries to connect two trees, one from the initial configuration and the other from the goal configuration. It has been used for path planning of ground vehicles in static environments. A reactive planning approach for navigating in dynamic environments has been reported by(\cite{tro2}). Reactive methods ensure only safety, but there is no guaranty on the optimality of the planned path. Optimal path-planning in 2D environments with dynamic obstacles is reported by \cite{ijrr8}. Accuracy of the optimality is limited by the polygonal approximation of the geodesic on the plane of the object surface. A preliminary work on motion planning in dynamic environments is due to \cite{icar02}, in that the authors proposed to generate a velocity obstacle using the non-linear velocity obstacle based motion planner. Further, a near time-optimal path planning strategy, namely, Dynamic Obstacle Velocity Space (DOVS) suitable for dynamic environments, has been very recently discussed by \cite{ijrr3}. A rigorous study on the velocity obstacle approach for an environment comprising of both static and dynamic obstacles is presented in this work. This is a model-based motion planning and navigation algorithm. The incremental nature of this algorithm limits the possibility to generate a lesser cost path and, at the same time, takes enormous time to generate the path. 

Another recent work by \cite{Chen2017} used GVO and, for the first time, showed the possibility of combining RRT* (applicable to static environment) with GVO for a typical hybrid (static and dynamic) environment. Typically, non-holonomic robots have been considered here for navigation in dynamic environments. These results motivate us to look into other motion planning algorithms that are typically used in static environments. 

    We further study the classic yet state-of-the-art path planners applicable to static environments. RRT*  is a  classical work for motion planning in static environments discussed by \cite{ijrr12}, that was used by \cite{Chen2017}. One of the very recent single-query motion planning algorithm that solves the problem in an asymptotically optimal sense is called RRT$^{\mathbf{x}}$. It has been detailed by \cite{ijrr10}. It is specifically designed to deal with dynamic obstacles. A Bayesian network has been used by \cite{iros16} to estimate effective global robot localization. As a further improvement to RRT*, informed RRT* was introduced by \cite{iros17}. It reduces the search space compared to normal RRT* and therefore reduces the time to sample a collision-free path. A new addition to this is Batch Informed Trees (BIT*) proposed by \cite{icra19}. BIT* finds out the optimal euclidean path in minimum time among the existing state-of-the-art path-planning algorithms in a static environment. Therefore, extending the investigation in this direction gives a possibility of using a potential static environment path planner applicable to hybrid environments, having both static and dynamic obstacles. 
    
\section{Motion Planning in Hybrid Environment}

\subsection{Objective}
In this paper, we are interested in finding out an optimal motion planning algorithm that is capable of planning a path by the expense of the minimum time, and the planned path is also the least cost in comparison with other state-of-the-art motion planning techniques. As discussed in the previous section and is also shown in Figure~\ref{fig:01}, currently available motion planning algorithms that deal with dynamic obstacles guarantee the safety of the planned path only. The cost of the path has not been addressed to be an important parameter. To be specific, DOVS, or other velocity obstacle based approaches to ensure safe planning within the visibility horizon. On the other hand, BIT* plans the globally least-cost path in static environments. Time optimality and path cost, both are given importance in this paper. In this work, we aim to achieve the globally least-cost path for an environment with static and dynamic objects and, at the same time, ensure minimum time to compute the path. 

			\begin{figure*}[t]
	\begin{center}
		\begin{minipage}[t]{8.cm}
			\centering
			\includegraphics[height = 2.7 in, width= \linewidth]{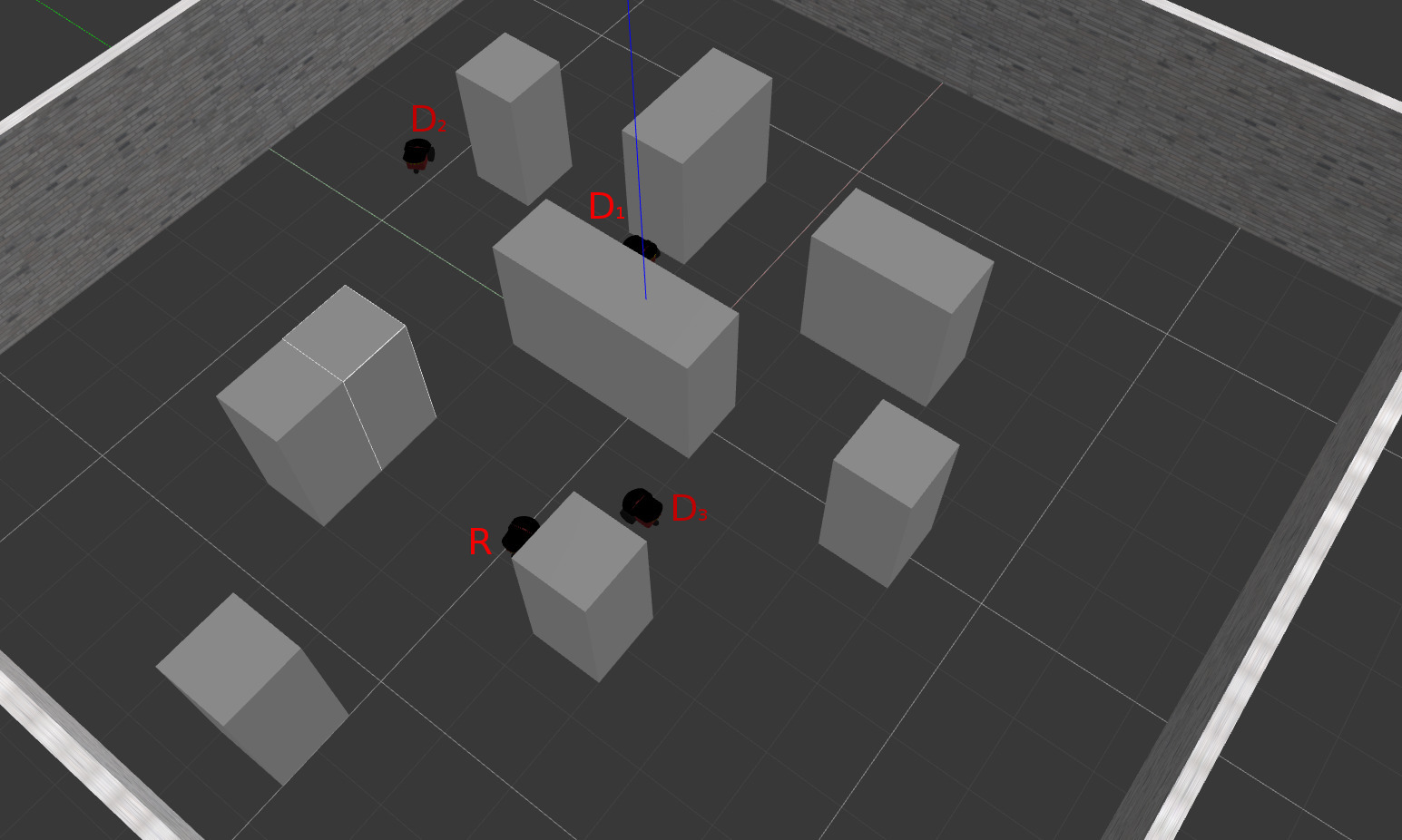}
		\end{minipage}
		\hspace{0cm}
		\begin{minipage}[t]{8.cm}
			\centering
			\includegraphics[height = 3 in, width= \linewidth]{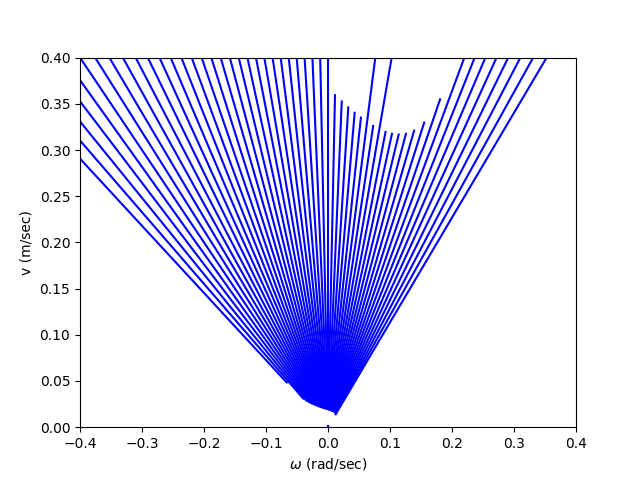}
		\end{minipage}
		\label{fig4}
	\end{center}
	\caption{(left) Environment representing three dynamic obstacles and a robot; (right) velocity space of the robot at the current instant is shown here.}
	\label{fig:02}
\end{figure*}
 In this section, we focus on the the state-of-the-art motion planning techniques that are used to build this work.
 
\subsection{Dynamic Obstacle Velocity Space}
Dynamic Obstacle Velocity Space (DOVS) was introduced to plan a safe and near time-optimal path primarily in the presence of one or more dynamic obstacle(s). The primary motive of this algorithm is to generate a set of forbidden velocities $(v, \omega)$ for the robot, based on its current velocity, and the motion parameters of the dynamic obstacle(s) within the visibility of the robot. This set of velocities then forms the Velocity Obstacle (VO), as shown in Figure~\ref{fig:02}. Any $(v,\omega)$ pair outside the DOV can be chosen for a safe navigation. In Figure~\ref{fig:02}, a simulated environment has been created that comprises static and dynamic obstacles. The DOV formed for the robot ($R$) at the current instant, due to the dynamic obstacle $D_3$ is shown at the right. As mentioned in the previous section, due to the incremental nature of this algorithm, at every sampling instant DOVS finds out a set of safe velocities for the robot, closest to the goal velocity $(v_g, \omega_g)$ outside the DOV. If it so happens that the current velocity of the robot is within the VO for the next sampling instant, a clothoid or anti-clothoid approach is followed, then to compute the set of motion commands to reach the new goal direction.  

Unlike the dynamic objects, static objects are not considered at every sampling instant. It is only when the distance between the robot and any static object is within a safe distance $D_{safe}$, defined by \cite{ijrr3}, the set of velocities that may lead to collision with the static obstacle(s), is included in the DOV. As static obstacles are not considered in the velocity obstacle of the robot until any static obstacle is within $D_{safe}$, this leaves a chance of collision for a robot with a static obstacle if the static obstacle was previously obscured by the dynamic obstacle and was situated close enough from with respect to the dynamic obstacle.

\subsection{Batch Informed Tree (BIT*)}
BIT* is a path planning algorithm that uses a sampling-based approach, in conjunction with graph-based planning, related to occupancy in the state space. It exploits a graph-like structure by forming a spanning tree. The tree of states is formed by sampling in the free space in the robot's state space in a given planning problem. The sampling mechanism in BIT* is much better directed than its predecessor sampling-based planners like RRT, RRT*, and informed RRT* by \cite{iros17}. BIT* asymptotically converges almost surely to the optimal solution and finds a probabilistically complete path. It means that for a given path cost if it finds a collision-free path that will be optimal and the probability that a solution can be found tends to 1 as more time is spent. This can also be expressed as given by \cite{icra19},
\begin{equation}
P\bigg(\limsup\limits_{q\rightarrow \infty} c_{\text{best},q}^{\text{BIT}^*} = s^*\bigg) = 1
\end{equation}
where, $c_{\text{best},q}^{\text{BIT}*}$ is the cost of the best solution $s^*$ found by BIT* from $q$ samples.
Based on the work by \cite{Chen2017}, and upon studying the BIT*, we realize that BIT* itself can be applied to dynamic scenarios, by incorporating few additional steps, proposed in the next section.    	 

\section{Propose Algorithm}
\begin{figure}[t]
	\begin{center}
		\includegraphics[width=8.4cm]{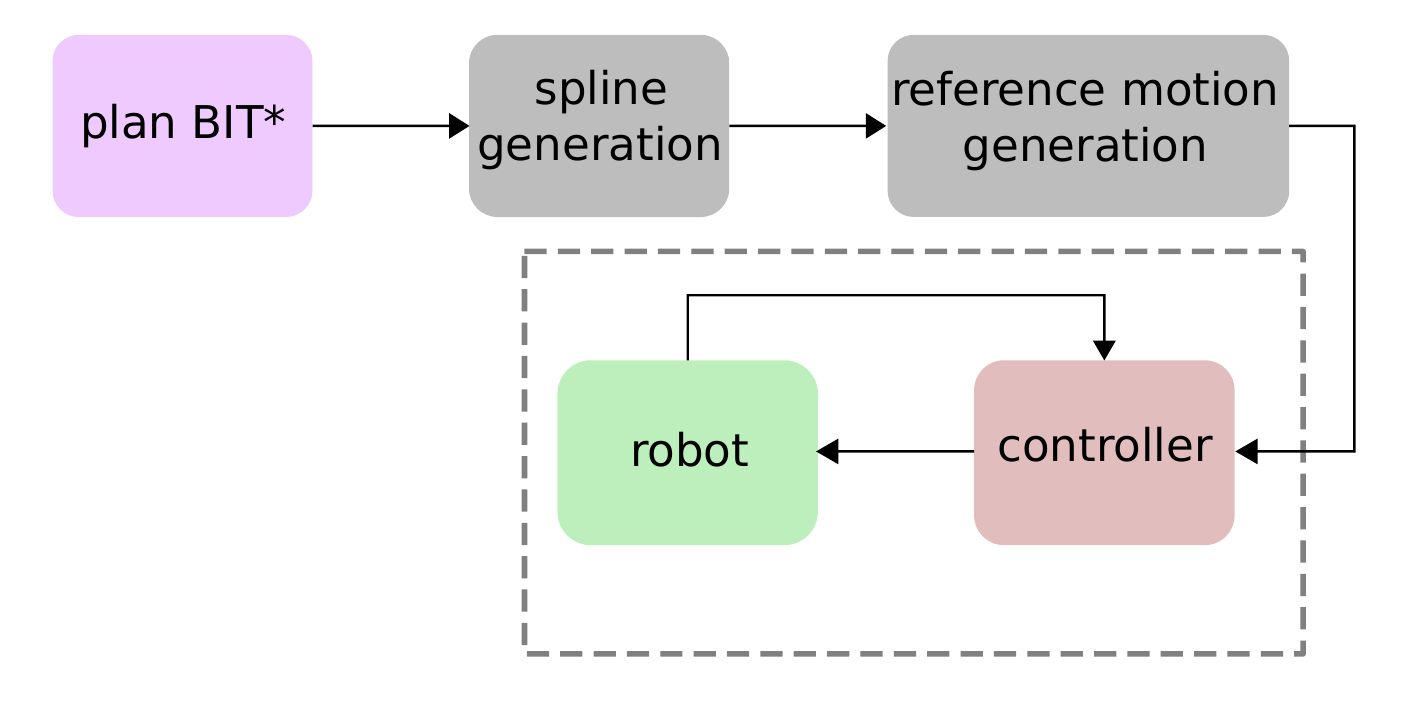}    
		\caption{motion planning block diagram} 
		\label{fig:03}
	\end{center}
\end{figure}

If the robot comes across a dynamic obstacle in its way to the goal, a specialized algorithm needs to be employed to re-plan a collision-free path. In this situation, one of the objectives is to minimize the re-planning time. This is due to two reasons, (a) while traveling along the previously planned path, the longer it takes to plan a collision-free path, the more the robot would require to travel to align with the newly planned path, (b) if the robot is stopped for a long while re-planning, keeps it far from the reference virtual robot in space and time. For (a), it does not serve any good towards achieving an optimal cost planned path. For (b), this is also not a very good planning technique as the robot lags way behind the reference robot position. 
Therefore, in this section, we propose a novel re-planning technique that requires the same amount of time that is necessary to plan a path in the presence of static obstacle(s). The entire motion planning algorithm proposed in this section is presented by a block diagram shown in Figure~\ref{fig:03}.

As mentioned in Section 3.3, BIT* is used for path planning in the re-planning phase (see Section 4.1). The path generated by this kind of sampling-based planning algorithms is a set of samples or nodes that can be joined individually to form a piecewise continuous segment that is the least cost in terms of the euclidean distance metric. The generated path may appear to be non-smooth in nature, and such kind of path does not qualify to generate motion on it. Therefore, to develop a complete motion planning algorithm, we need to generate a smooth and continuous path based on the pre-planned path. 

A number of options exist to interpolate the piecewise continuous segments obtained from BIT*. Cubic spline basis function is a good option for interpolation as it generates a stable continuous function and does not result in overfitting. Therefore, we use a cubic spline basis to generate a smooth and continuous path. For motion planning, we consider non-holonomic differential drive robots in this case. According to Figure~\ref{fig:03}, the next task is to generate the reference motion of the robot, so that the desired states of the robot along the path are generated. Once the continuous path is generated, based on the curvature of the path and the maximum allowable linear and angular velocity constraints of the vehicle, we compute the desired motion parameters of the robot.
Ideally, the robot should exactly match the reference states while moving along the planned path. We used the control law developed by \cite{mcc} for our case to generate the control that the robot should follow with.  

\subsection{Online Re-planning} 
The steps for re-planning is provided in Algorithm 1. When the robot encounters a new dynamic obstacle that is potentially going to lead to a collision, it needs to find a new collision free path. Let us consider the static map of the environment is known, i.e. the position of the static obstacles are known apriori. We assume that the motion parameters of the dynamic obstacles is known as well. It is still a challenging task to find an optimal path that ensures the optimality constraints, the matter of concern, mentioned in Section 3. 

\begin{algorithm}[t]
	\KwIn{source $S$, goal $G$, obstacle(s), robot and obstacle motion parameters}
	\KwOut{segmented path}
	initialization: $\mathcal{P}$ = $\emptyset$ \tcp{set of planned path,} $\mathcal{V} = \emptyset $  \tcp{set of virtual obstacles}
	
	\While{$(future$ $collision)$}{
		compute $\mathbf{x}$;  \tcp{point of collision}
		compute $v$; \tcp{ static object correspond to dynamic obstacle in space-time} 
		$p$ = plan BIT* \tcp{current planned path}
		$P = P\cup p$  \tcp{}
		\eIf{$( !collision$ $free)$}{
			$\mathcal{V} = \mathcal{V} \cup v$ \tcp{append new virtual obstacle}
			continue;
		}
		{$p*=p$ \tcp{optimal collision free path}
			break;}
	}
	\Return{p*}
	\caption{Sequential BIT*}
\end{algorithm}

Whenever, it is estimated that due to the relative motion between the robot and a dynamic obstacle, there is a future collision, we propose to occupy that region of collision by creating a virtually static obstacle, as shown in line 9, Algorithm 1. The size of this virtual obstacle is considered to be a square with side length twice of the total size of the obstacle, and the robot. This virtual static object is a representation of the dynamic object in space at that time.The re-planning is performed then by using BIT* algorithm, as shown in line 6, Algorithm 1. Based on the design of the algorithm, when, the first re-planned path is found the robot drives from the previously planned path to the newly planned one. The process of moving from one path to the other is done by first accelerating the robot about its axis of rotation up to $\omega_g{max}$, then decelerate it towards zero. It follows then by accelerating along the direction of translation on the newly planned path. Based on the reference motion generated on the newly planned path, if it so happens that the there exists a future collision with the dynamic obstacle, we further append another virtual static obstacle with the same size. The center of this virtual obstacle is placed at the point of collision on the currently planned path. 
Upon testing for multiple occasions, we find that we need to plan three paths on an average to find a collision free path and we do perform it by using parallel processing which effectively takes the same amount of time that BIT* requires to plan a path.

\section{Experiments}
In this section, we perform a number of simulation experiments to show that the proposed algorithm stands out with respect to the state-of-the-art based on a number of metrics mentioned below. The high level simulations of the dynamical system has been performed using GAZEBO and ROS.  Typically, the environment consists of both static and dynamic obstacles. Unlike, the point robot governed by only kinematic equations, here we have considered a Pioneer P3-DX robot for the simulation along with the dynamic constraints of it. The experimental environment is rectangular in size with a size of $(15\times 11)m^2$ area. Motion parameters of the robot and the dynamic obstacles are given in Table~\ref{tab:01}.

\begin{table}[t]
	\centering
	\caption{Robot and Obstacle(s) parameters}
	\begin{tabular}{|c||c|c|c|c|}
		\hline
		\hline
		parameter & $R$  & $D_1$ & $D_2$ & $D_3$ \\
		\hline
		$v (m/s)$ & 0.4  & 0.12 & 0.12 & 0.2\\
		\hline
		$\omega (rad/s)$ & 0.4 & 0 & -0.14 & 0 \\
		\hline
		$a_{max} (m/s^2)$ & 0.4 & - & - & - \\
		\hline
		$\alpha_{max} (rad/s^2)$ & 1 & - & - & - \\
		\hline\hline
	\end{tabular}
	\label{tab:01}
\end{table}

Although, DOVS is an incremental algorithm, but due to real time performance requirement of the path planner, we need to obtain the run time of each algorithm, as a performance metric.

Then, we perform a failure analysis which typically shows that the principle of computing $D_{safe}$ for DOVS, can cause a mobile robot leading in a collision before it is stopped. In addition to that, introducing multiple dynamic objects in the environment entangles the path planning process, thereby finding a collision free path becomes difficult. The detailed analysis of the simulation experiments follows in the next section.

\section{Results and Discussion}
To be able to compare the performance of the two algorithms, we chose (a) total path length, (b) total run time, and (c) time taken to reach the goal, as the metrics. Dynamic obstacles are commanded to run in with linear and circular trajectory.

\begin{figure*}[t]
	\begin{minipage}{.332\textwidth}
		\includegraphics[width=\linewidth]{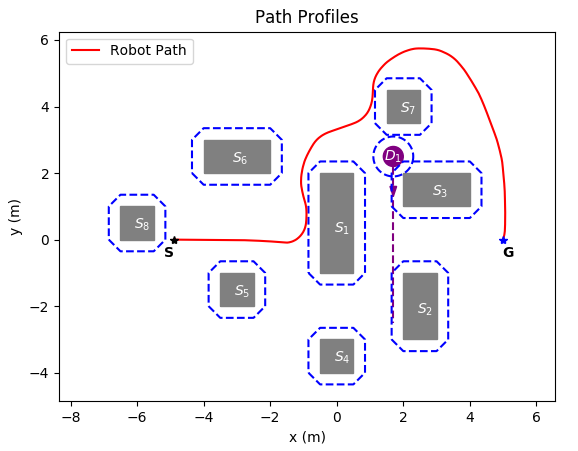}
	\end{minipage} \quad
	\begin{minipage}{.324\textwidth}
		\includegraphics[width=\linewidth]{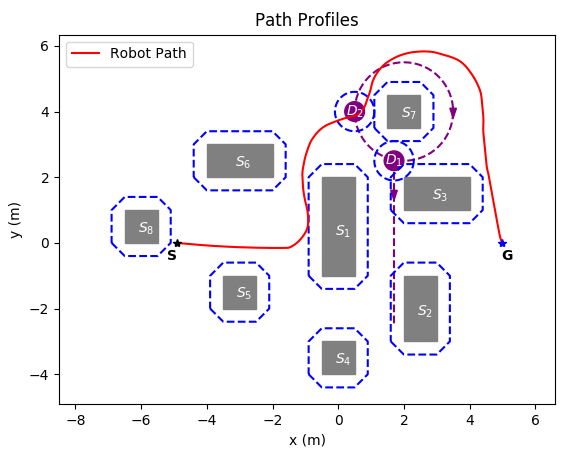}
	\end{minipage}\quad
	\begin{minipage}{.33\textwidth}
		\includegraphics[width=\linewidth]{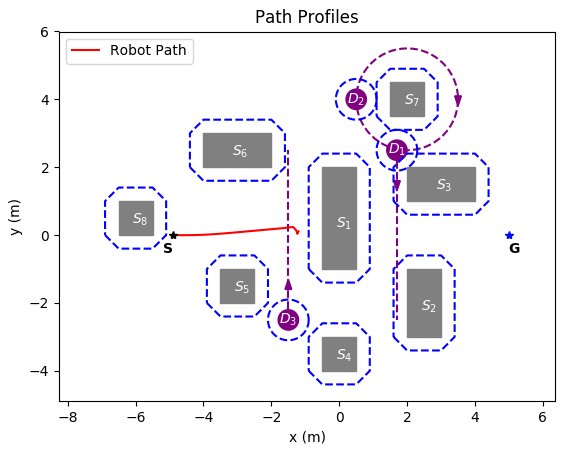}
	\end{minipage}\\
	
	\begin{minipage}{.33\textwidth}
		\includegraphics[width=\linewidth]{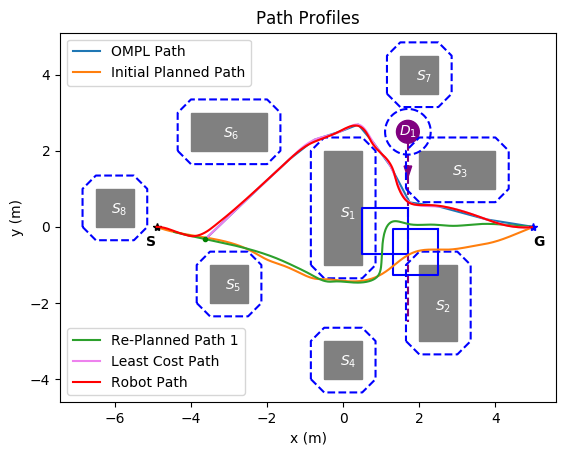}
	\end{minipage} \quad
	\begin{minipage}{.324\textwidth}
		\includegraphics[width=\linewidth]{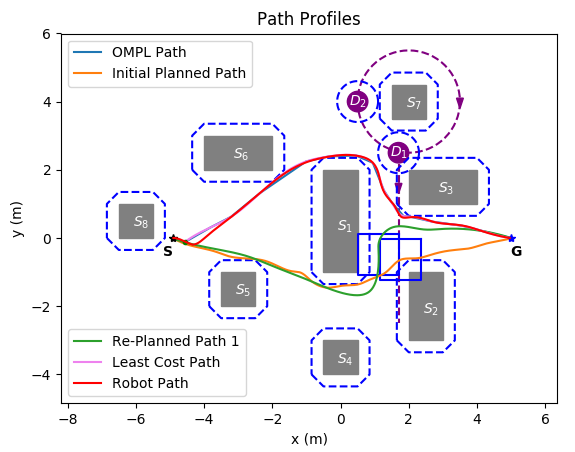}
	\end{minipage}\quad
	\begin{minipage}{.33\textwidth}
		\includegraphics[width=\linewidth]{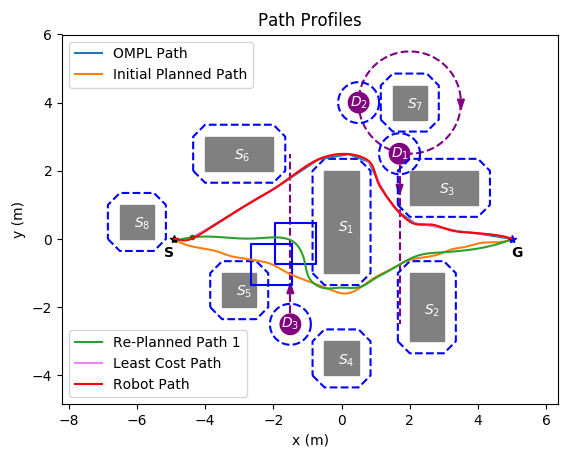}
	\end{minipage}
	\caption{(top row) present the DOVS simulation experimental results; (bottom row) shows the experimental results for sequential BIT* algorithm; both static $S_i$ as well as dynamic obstacle(s) $D_i$ are introduced in the simulation environment; three different environment setting have been used comprising of one, two and three dynamic obstacle(s) respectively; DOVS in (top left) requires much longer path (red) to reach the goal, compared with sequential BIT* in (bottom left) as shown in (red); DOVS in (top right) terminates due to the robot leading to a collision with dynamic obstacle $D_1$, whereas BIT* (bottom right) plans a collision free maneuverable path even in presence of three dynamic obstacles.}
	\label{fig:04}
\end{figure*}

Graphical results have been used to clearly demonstrate the paths generated by the two algorithms of central interest in this paper namely DOVS and the proposed sequential BIT*. These results enable us to qualitatively identify the best out of the two algorithms. In Figure~\ref{fig:04}, the results in the top row correspond to the DOVS algorithm in presence of one (a), two (b), and three dynamic obstacles, whereas, the bottom row represents the results of the proposed sequential BIT* algorithm. For one and two obstacle(s) cases, DOVS is able to complete the task by reaching the goal whereas, upon inclusion of the third dynamic obstacle, it crashes as shown in the (top right) in Figure~\ref{fig:04}. Sequential BIT* on the other hand, is able to reach the goal by following an almost similar path (qualitatively) in the same environment, that it generated for the two previous environment configurations. In all three cases, it is clearly observed that sequential BIT* plans much tighter path than DOVS. Blue rectangular regions in Figure~\ref{fig:04}, represent the virtual static obstacles as mentioned in Algorithm 1. Two of such rectangles were enough to ensure collision avoidance and the optimal planned path was obtained, upon considering the region occupied by these two rectangles as static object.

Based on the prior mentioned metrics, a quantitative performance comparison of these two algorithms are provided in Table~\ref{tab:02} and Table~\ref{tab:03} respectively. Table~\ref{tab:02} reports that the length of the total path traversed by the robot for the first two environment configurations are at least $2/3$rd larger than that of the sequential BIT* provided in Table~\ref{tab:03}. BIT*, being the sampling based motion planning algorithm, the planned path may vary each time, based on the samples drawn from the generating distribution. Therefore, to obtain a reliable path length for each environment, we run the proposed motion planning algorithm for $n$ times and take the mean over the obtained {\em path length}. The reported statistics is the mean of the outcome of 30 different runs of the experiment. The {\em time to goal} is also calculated in the same manner.

\subsection{Future work}
The results reported in this paper leaves us with a number of research possibilities. A robust motion planner is necessary that will relax the requirement of sequential nature of the currently proposed algorithm in this paper. It will enable a mobile robot to follow a better cost, and more reliable path in dynamic environment. To be precise, such a planner should address the spatio-temporal occupancy of dynamic object(s) around the robot more tightly such that typical static environment motion planner can be found very effective. 

\begin{table}[t]
	\centering
	\caption{DOVS Performance Statistics}
	\begin{tabular}{|c||c|c|c|}
		\hline
		\hline
		\begin{tabular}{@{}c@{}} Dynamic \\ obstacle $(\#)$ \end{tabular}
		& path length (m)  &
		\begin{tabular}{@{}c@{}}path generation \\ time (s) \end{tabular}&	\begin{tabular}{@{}c@{}} time to \\ goal (s) \end{tabular} 	  \\
		\hline
		1 & 18.762  & 0.25188 & 50.1 \\
		\hline
		2 & 18.4852  & 0.207315 & 56.006\\
		\hline
		3 &  \multicolumn{3}{c|}{crashed} \\
		\hline\hline
	\end{tabular}
	\label{tab:02}
\end{table}

\begin{table}[t]
	\centering
	\caption{BIT* Performance Statistics}
	\begin{tabular}{|c||c|c|c|}
		\hline
		\hline
		\begin{tabular}{@{}c@{}} Dynamic \\ obstacle $(\#)$ \end{tabular}
		& path length (m)  &
		\begin{tabular}{@{}c@{}}path generation \\ time (s) \end{tabular}&	\begin{tabular}{@{}c@{}} time to \\ goal (s) \end{tabular} 	  \\
		\hline
		1 & 12.076  & 0.1841 & 40.766 \\
		\hline
		2 & 12.003  & 0.1716 & 40.218\\
		\hline
		3 &  12.115 & 0.1835 & 40.655 \\
		\hline\hline
	\end{tabular}
	\label{tab:03}
\end{table}

\section{Conclusion}
In this paper, we proposed a novel anytime motion planning algorithm for a mobile robot in presence of one or more dynamic obstacle(s). Upon thorough investigation, based on planned path cost, time to plan and safety factor, we selected DOVS and BIT* for further study. Based on, multiple simulations and corresponding qualitative as well as quantitative results, it is shown that {\em Sequential BIT*} outperforms DOVS by large margin, in terms of the {\em path length}, {\em path generation time}, and {\em time to goal}. The impressive results of this work leaves with further scope of improvements towards achieving a lesser path cost as well as generating a tighter representation of dynamic obstacle in space-time.

\bibliographystyle{unsrt}  
\bibliography{ifacconf.bib}

%
%
%
%

\end{document}